\documentclass{article}%
\usepackage{amsmath}
\usepackage{amsfonts}
\usepackage{amssymb}
\usepackage[pdftex]{graphicx}%
\providecommand{\U}[1]{\protect\rule{.1in}{.1in}}
\pdfoutput=1

\begin{document}

\title{From Entropic Dynamics to Quantum Theory\thanks{Extended and corrected version
of a paper presented at MaxEnt 2009, the 29th International Workshop on
Bayesian Inference and Maximum Entropy Methods in Science and Engineering
(July 5-10, 2009, Oxford, Mississippi, USA).}}
\author{Ariel Caticha\\{\small Department of Physics, University at Albany-SUNY, }\\{\small Albany, NY 12222, USA. (ariel@albany.edu)}}
\date{}
\maketitle

\begin{abstract}
Non-relativistic quantum theory is derived from information codified into an
appropriate statistical model. The basic assumption is that there is an
irreducible uncertainty in the location of particles: positions constitute a
configuration space and the corresponding probability distributions constitute
a statistical manifold. The dynamics follows from a principle of inference,
the method of Maximum Entropy. The concept of time is introduced as a
convenient way to keep track of change. A welcome feature is that the entropic
dynamics notion of time incorporates a natural distinction between past and
future. The statistical manifold is assumed to be a dynamical entity: its
curved and evolving geometry determines the evolution of the particles which,
in their turn, react back and determine the evolution of the geometry.
Imposing that the dynamics conserve energy leads to the Schroedinger equation
and to a natural explanation of its linearity, its unitarity, and of the role
of complex numbers. The phase of the wave function is explained as a feature
of purely statistical origin. There is a quantum analogue to the gravitational
equivalence principle.

\end{abstract}

\section{Introduction}

Our subject has been very succinctly stated by Jaynes: \textquotedblleft Our
present QM formalism is a peculiar mixture describing in part realities in
Nature, in part incomplete human information about Nature---all scrambled up
by Heisenberg and Bohr into an omelette that nobody has seen how to
unscramble.\textquotedblright\ \cite{Jaynes 90}\ He also understood where to
start looking: \textquotedblleft We suggest that the proper tool for
incorporating human information into science is simply probability
theory---not the currently taught `random variable' kind, but the original
`logical inference' kind of James Bernoulli and Laplace\textquotedblright%
\ which he explains \textquotedblleft is often called Bayesian
inference\textquotedblright\ and is \textquotedblleft supplemented by the
notion of information entropy\textquotedblright.\ Bohr, Heisenberg and other
founders of quantum theory might have agreed. They were keenly aware of the
epistemological and pragmatic elements in quantum mechanics (see e.g.,
\cite{Stapp 72}) but they wrote at a time when the language and tools of
quantitative epistemology had not yet been sufficiently developed.

Our goal is to derive quantum theory as an example of entropic inference. A
central feature is the privileged role we assign to position over and above
all other observables. Position is, strictly, the only \emph{observable}. This
is one important difference from other approaches that also emphasize notions
of information (see \emph{e.g.}, \cite{Wootters 81}-\cite{Goyal 10}). The
theory has formal similarities with Nelson's stochastic mechanics \cite{Nelson
66}-\cite{Smolin 06}, but there are important conceptual differences.
Stochastic mechanics operates at the ontological level; its goal is a
realistic interpretation of quantum theory as arising from a deeper, possibly
non-local, but essentially classical \textquotedblleft
reality\textquotedblright. In contrast, the entropic dynamics advocated here
operates almost completely at the epistemological level.

The statistical manifold associated to the configuration space is introduced
in section 2. The basic dynamical question --- the probability of a small step
--- is answered in section 3 using the method of maximum entropy \cite{Caticha
08}. (Earlier versions of related ideas were discussed in \cite{Caticha
00}-\cite{Caticha Cafaro 07}.) The introduction of time as a device to keep
track of the accumulation of small changes is discussed in section 4 and the
Schroedinger equation is derived in section 5. An important new element is
that the geometry of the statistical manifold is not a fixed background but a
dynamical entity. This leads to a certain similarity with the theory of
general relativity including a new quantum equivalence principle (section 6).
Our conclusions are summarized in section 7.

\section{The statistical model}

We assume the usual configuration space $\mathcal{X}$ for a single particle,
namely, a flat three-dimensional space. We further assume --- and this is the
crucial new element --- that there is a very small but ultimately irreducible
uncertainty in the location of the particle. The particle cannot be localized;
at best we can specify its expected position $x^{a}=\langle y^{a}\rangle$ and
give a probability distribution $p(y|x)$. Thus, when we say that the particle
is at $x$ what we mean is that the true but unknown position $y$ is somewhere
in the vicinity of $x$ with probability $p(y|x)$.

To each point\ $x\in\mathcal{X}$ there corresponds a probability distribution
$p(y|x)$ and the set of these distributions is also a three-dimensional
manifold --- a \emph{statistical} manifold which we call $\mathcal{M}$. The
same label $x$ is used to denote points in $\mathcal{X}$ and the corresponding
points in $\mathcal{M}$. Points in $\mathcal{M}$ are not structureless dots
but probability distributions.

The origin of the irreducible uncertainty is left unspecified. Indeed, the
arguments below turn out to be remarkably independent of the particular choice
of the functional form of $p(y|x)$. Nevertheless, for the sake of clarity it
is convenient to choose a specific statistical model. We can reasonably assume
that the distributions $p(y|x)$ arise as the result of unknown microscopic
influences, in which case, general arguments such as the central limit theorem
lead us to expect that for a very wide variety of microscopic conditions the
plausible distributions are Gaussians. We further assume that the Gaussians
are spherically symmetric with a small non-uniform variance expressed as a
small constant $\sigma^{2}$ modulated by a positive scalar field $\Phi(x)$,
\begin{equation}
p(y|x)=\frac{\Phi^{3/2}(x)}{(2\pi\sigma^{2})^{3/2}}\,\exp\left[  -\frac{1}%
{2}\Phi(x)\gamma_{ab}(y^{a}-x^{a})(y^{b}-x^{b})\right]  ,\label{Gaussian}%
\end{equation}
where
\begin{equation}
\gamma_{ab}=\frac{\delta_{ab}}{\sigma^{2}}~\label{gamma}%
\end{equation}
is the metric in $\mathcal{X}$.

While the configuration space $\mathcal{X}$ is flat, the statistical manifold
$\mathcal{M}$ turns out, in general, to be curved. $\mathcal{M}$ inherits its
unique geometry from the distributions $p(y|x)$. The distance, $d\ell
^{2}=g_{ab}\,dx^{a}dx^{b}$, between $x$ and $x+dx$, or better, between
$p(y|x)$ and $p(y|x+dx)$, is given by the information metric \cite{Caticha
08}\cite{Amari 00}\cite{Cencov81},
\begin{equation}
g_{ab}=\int dy\,p(y|x)\frac{\partial\log p(y|x)}{\partial x^{a}}\frac
{\partial\log p(y|x)}{\partial x^{b}}~.\label{info metric a}%
\end{equation}
Substituting (\ref{Gaussian}) and (\ref{gamma}) into (\ref{info metric a})
gives
\begin{equation}
g_{ab}(x)=\frac{\Phi}{\sigma^{2}}\delta_{ab}+\frac{3}{2\Phi^{2}}\partial
_{a}\Phi\partial_{b}\Phi~.
\end{equation}
We will be interested in situations where the intrinsic uncertainties are very
small. More precisely, when the change in $\Phi(x)$ over the support of
$p(y|x)$ is negligible, $|\partial\Phi/\Phi|^{2}\ll\Phi/\sigma^{2}$, the
metric simplifies considerably,
\begin{equation}
g_{ab}(x)\approx\frac{\Phi(x)}{\sigma^{2}}\delta_{ab}=\Phi(x)\gamma
_{ab}~,\label{info metric b}%
\end{equation}
and we see that $\Phi(x)$ plays the role of a conformal factor.

For future reference, the entropy of $p(y|x)$ relative to the flat measure of
$\mathcal{X}$ is
\begin{equation}
S(x)=-\int dy\,p(y|x)\log\frac{p(y|x)}{\gamma^{1/2}}=\frac{3}{2}[1-\log
\frac{\Phi(x)}{2\pi}].\label{entropy a}%
\end{equation}
(The volume element in $\mathcal{X}$ is given by $dv=\gamma^{1/2}d^{3}x$ where
$\gamma=\det\gamma_{ab}$.)

The generalization to $N$ particles is straightforward. The $3N$-dimensional
configuration space $\mathcal{X}_{N}$ remains flat but it is no longer
isotropic. For example for $N=2$ particles its metric is
\begin{equation}
\gamma_{AB}=%
\begin{bmatrix}
\delta_{a_{1}b_{1}}/\sigma_{1}^{2} & 0\\
0 & \delta_{a_{2}b_{2}}/\sigma_{2}^{2}%
\end{bmatrix}
~.\label{gamma AB}%
\end{equation}
The position uncertainty is given by a Gaussian distribution $p(y|x)$ in $3N$
dimensions,
\begin{equation}
p(y|x)\propto\,\exp\left[  -\frac{1}{2}\Phi(x)\gamma_{AB}(y^{A}-x^{A}%
)(y^{B}-x^{B})\right]  ,
\end{equation}
where the index $A$ takes the values $1\ldots3N$. The statistical manifold
$\mathcal{M}_{N}$ has metric $g_{AB}(x)\approx\Phi(x)\gamma_{AB}$.

\section{Law without law: entropic dynamics}

The basic dynamical information is that changes from one state to another are
possible and do, in fact, happen. We do not explain why they happen but, given
the information that changes occur, we want to venture a guess about what
changes to expect. What gives this program some hope of success is the
assumption that large changes result from the accumulation of many small
changes. Therefore, our job separates into two parts, first we consider a
small change, and then we figure out how small changes add up.

Consider a single particle that moves from an initial position $x$ to an
unknown final position $x^{\prime}$. (The generalization to more particles is
immediate.) What can we say about $x^{\prime}$ when all we know is that it is
near $x$? Since $x$ and $x^{\prime}$ represent probability distributions we
use the method of maximum entropy (ME) \cite{Caticha 08}. As in all ME
problems success hinges on appropriate choices of entropy, prior distribution,
and constraints. Since neither the new $x^{\prime}$ nor the new microstate
$y^{\prime}$ are known, the relevant universe of discourse is
$\mathcal{X\times X}$ and what we want to find is the joint distribution
$P(x^{\prime},y^{\prime}|x)$ \cite{Rodriguez 98}. The appropriate (relative)
entropy is
\begin{equation}
\mathcal{S}[P,\pi]=-\int dx^{\prime}dy^{\prime}\,P(x^{\prime},y^{\prime
}|x)\log\frac{P(x^{\prime},y^{\prime}|x)}{\pi(x^{\prime},y^{\prime}%
)}~.\label{Sppi}%
\end{equation}
The relevant information is introduced through the prior $\pi(x^{\prime
},y^{\prime})$ and the constraints that specify the family of acceptable
posteriors $P(x^{\prime},y^{\prime}|x)$. Consider the prior, $\pi(x^{\prime
},y^{\prime})=\pi(x^{\prime})\pi(y^{\prime}|x^{\prime})$. Before the relation
between the variables $x^{\prime}$ and $y^{\prime}$ is known the state of
extreme ignorance is represented by a product, $\pi(x^{\prime},y^{\prime}%
)=\pi(x^{\prime})\pi(y^{\prime})$ --- knowledge of $x^{\prime}$ tells us
nothing about $y^{\prime}$ and vice versa --- and the probabilities
$\pi(y^{\prime})d^{3}y^{\prime}$ and $\pi(x^{\prime})d^{3}x^{\prime}$ are
uniform, that is, proportional to the respective volume elements $\gamma
^{1/2}d^{3}y^{\prime}$ and $\gamma^{1/2}d^{3}x^{\prime}$. Proportionality
constants are not essential here; we set $\pi(y^{\prime})=\gamma^{1/2}$ and
$\pi(x^{\prime})=\gamma^{1/2} $ and the prior is
\begin{equation}
\pi(x^{\prime},y^{\prime})=\gamma~.\label{prior}%
\end{equation}

Next consider the possible posteriors $P(x^{\prime},y^{\prime}|x)=P(x^{\prime
}|x)P(y^{\prime}|x^{\prime},x)$. Besides normalization we impose two
constraints. First we have the known relation between $x^{\prime} $ and
$y^{\prime}$: the particular functional form of $P(y^{\prime}|x^{\prime},x)$
is given by eq.(\ref{Gaussian}). Therefore,
\begin{equation}
P(x^{\prime},y^{\prime}|x)=P(x^{\prime}|x)p(y^{\prime}|x^{\prime
})~.\label{constraint p}%
\end{equation}
The second constraint concerns the factor $P(x^{\prime}|x)$: we know that
$x^{\prime}$ is only a short step away from $x$. Let $x^{\prime a}%
=x^{a}+\Delta x^{a}$. We require that the expectation
\begin{equation}
\left\langle \Delta\ell^{2}\right\rangle =\left\langle \gamma_{ab}\Delta
x^{a}\Delta x^{b}\right\rangle =\lambda^{2}(x)~\label{constraint lambda}%
\end{equation}
be some small but for now unspecified numerical value $\lambda^{2}(x)$ which
might perhaps depend on $x$. (Provided the steps are sufficiently short their
actual length is not particularly critical; they do not even have to be all of
the same length.)

Once prior and constraints have been specified the ME method takes over.
Substituting (\ref{prior}) and (\ref{constraint p}) into (\ref{Sppi}) gives
\begin{equation}
\mathcal{S}[P,\pi]=-\int dx^{\prime}\,P(x^{\prime}|x)\log\frac{P(x^{\prime
}|x)}{\gamma^{1/2}}+\int dx^{\prime}\,P(x^{\prime}|x)S(x^{\prime
})~,\label{Sppi b}%
\end{equation}
where $S(x)$ is given in eq.(\ref{entropy a}). Varying $P(x^{\prime}|x)$ to
maximize $\mathcal{S}[P,\pi]$ subject to (\ref{constraint lambda}) and
normalization gives
\begin{equation}
P(x^{\prime}|x)=\frac{1}{\zeta(x,\alpha)}e^{S(x^{\prime})-\frac{1}{2}%
\alpha(x)\Delta\ell^{2}},\label{Prob xp/x}%
\end{equation}
where $\zeta(x,\alpha)$ is the normalization constant, and the Lagrange
multiplier $\alpha(x)$ is determined from $\partial\log\zeta/\partial
\alpha=-\lambda^{2}/2$. We see that large values of $\alpha(x)$ clearly lead
to short steps.

$P(x^{\prime}|x)$\emph{\ gives the probability of a step from }$x$\emph{\ to
}$x^{\prime}$\emph{. It\ is the basic building block out of which all dynamics
is constructed.} It implements what Wheeler foresaw as \textquotedblleft law
without law\textquotedblright: \textquotedblleft that every law of physics,
pushed to the extreme, will be found statistical and approximate, not
mathematically perfect and precise.\textquotedblright\ \cite{Wheeler Zurek 83}

The most probable displacement $\Delta\bar{x}^{a}$ is that which maximizes the
scalar probability density. For large $\alpha(x)$ we expect $\Delta x^{a}$ to
be small. Then
\begin{equation}
0=\frac{\partial}{\partial x^{\prime a}}\left[  S(x^{\prime})-\frac{1}%
{2}\alpha(x)\gamma_{bc}\Delta x^{b}\Delta x^{c}\right]  _{\Delta x=\Delta
\bar{x}}=\partial_{a}S(x)-\alpha(x)\gamma_{ab}\Delta\bar{x}^{b}~
\end{equation}
so that the maximum occurs at
\begin{equation}
\Delta\bar{x}^{a}=\frac{1}{\alpha(x)}\gamma^{ab}\partial_{b}S(x)~,
\end{equation}
which shows that the particle tends to drift up the entropy gradient.
Expanding the exponent of (\ref{Prob xp/x}) about its maximum gives%
\begin{equation}
P(x^{\prime}|x)\approx\frac{1}{Z(x)}\exp\left[  -\frac{\alpha(x)}{2\sigma^{2}%
}\delta_{ab}\left(  \Delta x^{a}-\Delta\bar{x}^{a}\right)  \left(  \Delta
x^{b}-\Delta\bar{x}^{b}\right)  \right]  ,\label{Prob xp/x b}%
\end{equation}
with a new normalization $Z(x)$. The displacement $\Delta x^{a}$ can be
written as its expectation plus a \textquotedblleft
fluctuation\textquotedblright,%
\begin{equation}
\Delta x^{a}=\Delta\bar{x}^{a}+\Delta w^{a}~,\label{Delta x}%
\end{equation}
where
\begin{align}
\left\langle \Delta x^{a}\right\rangle  & =\Delta\bar{x}^{a}=\frac{\sigma^{2}%
}{\alpha(x)}\delta^{ab}\partial_{b}S(x)~,\label{ED drift}\\
\left\langle \Delta w^{a}\right\rangle  & =0\quad\text{and}\quad\left\langle
\Delta w^{a}\Delta w^{b}\right\rangle =\frac{\sigma^{2}}{\alpha(x)}\delta
^{ab}~.\label{ED fluctuations}%
\end{align}
We see that as $\alpha\rightarrow\infty$ the steps get smaller, $\Delta\bar
{x}^{a}\rightarrow0$ as $\alpha^{-1}$, but the fluctuations become dominant
because $\Delta w^{a}\rightarrow0$ only as $\alpha^{-1/2}$. This implies that
as $\alpha\rightarrow\infty$ the trajectory is continuous, but not
differentiable---just as in Brownian motion.

\section{Time}

To keep track of the accumulation of small changes we need to introduce a
notion of time. Our task here is to develop a model that includes (a)
something one might identify as an \textquotedblleft instant\textquotedblright%
, (b) a sense in which these instants might be \textquotedblleft
ordered\textquotedblright, and (c) a useful concept of \textquotedblleft
duration\textquotedblright\ measuring the interval between instants. This set
of concepts constitutes what we will call \textquotedblleft
time\textquotedblright\ in entropic dynamics. A welcome bonus is that there is
an intrinsic directionality from past to future instants; an arrow of time is
generated automatically and need not be externally imposed.

The foundation to any notion of time is dynamics. Given an initial position we
have some idea, given by $P(x^{\prime}|x)$, of what the next position might
be. For all steps after the first, however, we are uncertain about both the
initial $x$ and the final step $x^{\prime}$, which means we must deal with the
joint probability $P(x^{\prime},x)$. Using the product rule $P(x^{\prime
},x)=P(x^{\prime}|x)P(x)$ and integrating over $x$, we get
\begin{equation}
P(x^{\prime})=%
{\textstyle\int}
dx\,P(x^{\prime}|x)P(x)~.\label{CK a}%
\end{equation}
If $P(x)$ happens to be the probability of different values of $x$ \emph{at a
given instant of time }$t$, then it is tempting to interpret $P(x^{\prime})$
as the probability of values of $x^{\prime}$ at a later instant of time
$t^{\prime}=t+\Delta t$. Accordingly, we write $P(x)=\rho(x,t)$ and
$P(x^{\prime})=\rho(x^{\prime},t^{\prime})$ so that
\begin{equation}
\rho(x^{\prime},t^{\prime})=%
{\textstyle\int}
dx\,P(x^{\prime}|x)\rho(x,t)~.\label{CK b}%
\end{equation}
Nothing in the laws of probability forces this interpretation on us---it is an
independent assumption about what constitutes time in the model. \emph{We use
eq.(\ref{CK b}) to define what we mean by an instant}: if $\rho(x,t)$ refers
to an \textquotedblleft initial\textquotedblright\ instant, then we use
$\rho(x^{\prime},t^{\prime})$ to define what we mean by the \textquotedblleft
next\textquotedblright\ instant. Thus, eq.(\ref{CK b}) allows time to be
constructed, step by step, as a succession of instants.

Specifying the interval of time $\Delta t$ between successive instants amounts
to tuning the steps, or equivalently $\alpha(x)$, appropriately. To model a
\textquotedblleft Newtonian\textquotedblright\ time that flows
\textquotedblleft equably\textquotedblright\ everywhere, that is, at the same
rate at all places and times we define $\Delta t$ as being independent of $x$,
and such that every $\Delta t$ is as long as the previous one. Inspection of
the actual dynamics as given in eq.(\ref{Prob xp/x b}-\ref{ED fluctuations})
shows that this is achieved if we choose $\alpha(x)$ so that
\begin{equation}
\alpha(x)=\frac{\tau}{\Delta t}=\operatorname{constant}~.\label{alpha}%
\end{equation}
where $\tau$ is a constant introduced so that $\Delta t$ has units of time.

Thus, it is the equable flow of time that leads us to impose uniformity on the
step sizes $\lambda^{2}(x)$ and the corresponding multipliers $\alpha(x)$.
This completes the implementation of Newtonian time in entropic dynamics. In
the end, however, the only justification for any definition of duration is
that it simplifies the description of motion, and indeed, eqs.(\ref{Delta x}%
-\ref{ED fluctuations}) are simplified to
\begin{equation}
\Delta x^{a}=b^{a}(x)\Delta t+\Delta w^{a}\label{Delta x b}%
\end{equation}
where the drift velocity $b^{a}(x)$ and the fluctuation $\Delta w^{a}$ are
\begin{align}
b^{a}(x)  & =\frac{\sigma^{2}}{\tau}\,\delta^{ab}\partial_{b}%
S(x)~,\label{drift future}\\
\left\langle \Delta w^{a}\right\rangle  & =0\quad\text{and}\quad\left\langle
\Delta w^{a}\Delta w^{b}\right\rangle =\frac{\sigma^{2}}{\tau}\Delta
t\,\delta^{ab}~.\label{fluc}%
\end{align}

Equation (\ref{drift future}) gives the \emph{mean velocity to the future} or
\emph{future drift},
\begin{equation}
b^{a}(x)=\lim_{\Delta t\rightarrow0^{+}}\frac{\left\langle x^{a}(t+\Delta
t)\right\rangle _{x(t)}-x^{a}(t)}{\Delta t}=\lim_{\Delta t\rightarrow0^{+}%
}\frac{1}{\Delta t}%
{\textstyle\int}
dx^{\prime}\,P(x^{\prime}|x)\Delta x^{a}\label{drift future b}%
\end{equation}
where $x=x(t)$, $x^{\prime}=x(t+\Delta t)$, and $\Delta x^{a}=x^{\prime
a}-x^{a}$. The expectation in (\ref{drift future b}) is conditional on the
earlier position $x=x(t)$. One can also define a \emph{mean velocity from the
past} or \emph{past drift},
\begin{equation}
b_{\ast}^{a}(x)=\lim_{\Delta t\rightarrow0^{+}}\frac{x^{a}(t)-\left\langle
x^{a}(t-\Delta t)\right\rangle _{x(t)}}{\Delta t}\label{drift past}%
\end{equation}
where the expectation is conditional on the later position $x=x(t)$. Shifting
the time by $\Delta t$, $b_{\ast}^{a}$ can be equivalently written as
\begin{equation}
b_{\ast}^{a}(x^{\prime})=\lim_{\Delta t\rightarrow0^{+}}\frac{x^{a}(t+\Delta
t)-\left\langle x^{a}(t)\right\rangle _{x(t+\Delta t)}}{\Delta t}=\lim_{\Delta
t\rightarrow0^{+}}\frac{1}{\Delta t}%
{\textstyle\int}
dx\,P(x|x^{\prime})\Delta x^{a}~,\label{drift past b}%
\end{equation}
with the same definition of $\Delta x^{a}$ as in eq.(\ref{drift future b}).

The two mean velocities, to the future $b^{a}$, and from the past $b_{\ast
}^{a}$, need not coincide. The connection between them is well known
\cite{Nelson 66}\cite{Nelson 85},%

\begin{equation}
b_{\ast}^{a}(x,t)=b^{a}(x)-\frac{\sigma^{2}}{\tau}\partial^{a}\log
\rho(x,t)~,\label{drift past c}%
\end{equation}
where $\partial^{a}=\delta^{ab}\partial_{b}$ and $\rho(x,t)=P(x)$. What might
not be widely appreciated is that eq.(\ref{drift past c}) is a straightforward
consequence of Bayes' theorem,
\begin{equation}
P(x|x^{\prime})=\frac{P(x)}{P(x^{\prime})}P(x^{\prime}|x)~.\label{bt1}%
\end{equation}
(For a related idea see \cite{Jaynes 89}.) The proof of eq.(\ref{drift past c}%
) is lengthy but straightforward; the crucial step is to Taylor expand
$P(x^{\prime})$ about $x$ in (\ref{bt1}) to get
\begin{equation}
P(x|x^{\prime})=\left[  1-(\partial_{b}\log\rho)\,\Delta x^{b}+\ldots\right]
P(x^{\prime}|x)\,,\label{bt2}%
\end{equation}
which accounts for the $\partial\log\rho$ term in eq.(\ref{drift past c}).

The fact that $b^{a}\neq$ $b_{\ast}^{a}$ is very significant because it
signals an asymmetry in time. The arrow of time, constitutes a puzzle that has
plagued physics ever since Boltzmann. The standard formulation of the problem
is that the laws of nature are symmetric under time reversal but everything
else in nature indicates a clear asymmetry. How can that be?

The notion that the laws of physics might be rules for the manipulation of
information and not necessarily laws of nature offers a new perspective on
this old puzzle. Note that entropic dynamics does not assume any underlying
laws of nature --- whether they be symmetric or not --- and it makes no
attempt to explain the asymmetry between past and future. The asymmetry is the
inevitable consequence of entropic inference. From the point of view of
entropic dynamics the challenge does not consist in explaining the arrow of
time; on the contrary, it is the reversibility of the laws of physics that
demands an explanation. Indeed, \emph{time itself always remains intrinsically
directional} even when the derived laws of physics turn out to be fully reversible.

\section{The Schroedinger equation}

Time has been introduced as a useful device to keep track of the accumulation
of small changes. The technique to do this is well known from diffusion
theory. The equation of evolution for the distribution $\rho(x,t)$, derived
from eq.(\ref{CK b}) together with (\ref{Delta x b})-(\ref{fluc}), is the
Fokker-Planck equation, \cite{Nelson 85}\cite{Chandrasekhar 43}
\begin{equation}
\partial_{t}\rho=-\partial_{a}\left(  b^{a}\rho\right)  +\frac{\sigma^{2}%
}{2\tau}\nabla^{2}\rho~,\label{FP forward}%
\end{equation}
where $\delta^{ab}\partial_{a}\partial_{b}=\nabla^{2}$. Using
eq.(\ref{drift past c}) it can be expressed in terms of $b_{\ast}^{a}$,
\begin{equation}
\partial_{t}\rho=-\partial_{a}\left(  b_{\ast}^{a}\rho\right)  -\frac
{\sigma^{2}}{2\tau}\nabla^{2}\rho~.\label{FP backward}%
\end{equation}
Adding eqs.(\ref{FP forward}) and (\ref{FP backward}) gives a continuity
equation
\begin{equation}
\partial_{t}\rho=-\partial_{a}\left(  v^{a}\rho\right)  \quad\text{where}\quad
v^{a}\overset{\text{def}}{=}\frac{1}{2}\left(  b^{a}+b_{\ast}^{a}\right)
~,\label{continuity}%
\end{equation}
where $v^{a}$ is interpreted as the \emph{velocity of the probability flow} or
the \emph{current velocity}. On the other hand, eq.(\ref{drift past c}) gives
yet another \textquotedblleft velocity\textquotedblright%

\begin{equation}
u^{a}\overset{\text{def}}{=}\frac{1}{2}\left(  b_{\ast}^{a}-b^{a}\right)
=-\frac{\sigma^{2}}{2\tau}\partial^{a}\log\rho~.\label{osmo}%
\end{equation}
Its interpretation follows from $v^{a}=b^{a}+u^{a}$. The drift $b^{a}$
represents the tendency of the probability $\rho$ to flow up the entropy
gradient while $u^{a}$ represents its tendency to slide down the density
gradient. The situation is analogous to Brownian motion where the drift
velocity is the response to the gradient of an external potential, while
$u^{a}$ is a response to the concentration gradient --- the
so-called\emph{\ osmotic force}. Accordingly, $u^{a}$ is called the
\emph{osmotic velocity}. Its contribution to the probability flow is the
actual diffusion current,
\begin{equation}
\rho u^{a}=-\frac{\sigma^{2}}{2\tau}\partial^{a}\rho~,
\end{equation}
which can be recognized as Fick's law, with a diffusion coefficient given by
$\sigma^{2}/2\tau$.

Since both the drift $b^{a}$ and the osmotic velocity $u^{a}$ are gradients,
it follows that $v^{a}=b^{a}+u^{a}$ is a gradient too,
\begin{equation}
v^{a}=\frac{\sigma^{2}}{\tau}\,\partial^{a}\phi\quad\text{where}\quad
\phi(x,t)=S(x)-\log\rho^{1/2}(x,t)~.\label{curr}%
\end{equation}

Eqs.(\ref{FP forward}-\ref{curr}) provide the complete answer to the problem
as originally posed: what is the expected evolution of a particle afflicted by
the intrinsic uncertainties described by $p(y|x)$ in eq.(\ref{Gaussian})?
Unfortunately, we can see that the dynamics described by eqs.(\ref{FP forward}%
-\ref{curr}) is not quantum mechanics; it is just diffusion. Indeed, in order
to construct a wave function, $\Psi=\rho^{1/2}e^{i\phi}$, in addition to the
density $\rho$ we need a second degree of freedom, a phase $\phi$.

Note that the function $\phi(x,t)$ in eq.(\ref{curr}) does not (yet) qualify
as an independent degree of freedom because the entropy $S(x)$ --- or
equivalently the conformal factor $\Phi(x)$ --- is an externally prescribed
field. As long as the statistical manifold is a fixed static manifold there is
no logical room for additional degrees of freedom. To promote $\phi(x,t)$ to
an independent degree of freedom we are forced to allow the manifold
$\mathcal{M}$ itself to participate in the dynamics.

To specify the dynamics of the manifold we follow Nelson and assume that it is
\textquotedblleft conservative\textquotedblright\ \cite{Nelson 79}. Requiring
that some \textquotedblleft energy\textquotedblright\ be conserved may seem
natural in that it clearly represents physically relevant information but we
feel that it is an assumption that demands a deeper justification. Normally
energy is whatever happens to be dynamically conserved as a result of
invariance under translations in time. But our dynamics has not yet been fully
defined; what, then, is \textquotedblleft energy\textquotedblright\ and why
should it be conserved in the first place? This is a question we leave for the
future. At this early stage, for the purpose of deriving a non-relativistic
model, we just propose an intuitively reasonable conserved energy and proceed.

The energy is chosen to be a local functional that includes a term
representing a potential energy and includes terms in the velocities that are
invariant under time reversal and under rotations. Under time reversal,
$t\rightarrow-t$, we have $b^{a}\rightarrow-b_{\ast}^{a}\,$, $v^{a}%
\rightarrow-v^{a}\,$, $u^{a}\rightarrow u^{a}$. For low velocities this means
we need only include quadratic terms in the velocities, $v^{2}$ and $u^{2}$
\cite{Smolin 06}. The proposed energy functional is
\begin{equation}
E[\rho,v]=\int d^{3}x\,\rho(x,t)\left(  A\gamma_{ab}v^{a}v^{b}+B\gamma
_{ab}u^{a}u^{b}+V(x)\right)  ~,\label{energy a}%
\end{equation}
where $A$ and $B$ are constants. In order that $E$ have units of energy
$A/\sigma^{2}$ and $B/\sigma^{2}$ must have units of mass. Then
\begin{equation}
E[\rho,v]=\int d^{3}x\,\rho(x,t)\left(  \frac{1}{2}mv^{2}+\frac{1}{2}\mu
u^{2}+V(x)\right)  ~,\label{energy b}%
\end{equation}
where $m=2A/\sigma^{2}$ and $\mu=2B/\sigma^{2}$ will be referred to as the
\textquotedblleft mass\textquotedblright\ and the \textquotedblleft osmotic
mass\textquotedblright\ respectively. It is further convenient to combine the
constants $\tau$ and $A$ into yet a new constant $\eta$, which relates the
units of mass or energy with those of time,
\begin{equation}
\eta=\frac{2A}{\tau}\quad\text{so that}\quad\frac{\sigma^{2}}{\tau}=\frac
{\eta}{m}~.\label{eta}%
\end{equation}
Then the current and osmotic velocities, eqs.(\ref{curr}) and (\ref{osmo}),
are
\begin{equation}
mv_{a}=\eta\,\partial_{a}\phi\quad\text{and\quad}mu_{a}=\eta\partial_{a}%
\log\rho^{1/2}~,\label{curr osmo}%
\end{equation}
while (\ref{energy b}) becomes%
\begin{equation}
E=\int dx\,\rho\left(  \frac{\eta^{2}}{2m}(\partial_{a}\phi)^{2}+\frac{\mu
\eta^{2}}{8m^{2}}(\partial_{a}\log\rho)^{2}+V\right)  ~.\label{energy c}%
\end{equation}

Next we impose that the energy $E[\rho,\phi]$ be conserved. After some
manipulations involving integration by parts and the continuity equation,
\begin{equation}
\dot{\rho}=-\partial_{a}\left(  \rho v^{a}\right)  =-\frac{\eta}{m}%
\partial^{a}\left(  \rho\partial_{a}\phi\right)  =-\frac{\eta}{m}\left(
\partial^{a}\rho\partial_{a}\phi+\rho\nabla^{2}\phi\right)  ~,\label{SEa}%
\end{equation}
the time derivative $\dot{E}$ of (\ref{energy c}) is
\begin{equation}
\dot{E}=\int dx\,\dot{\rho}\left[  \eta\dot{\phi}+\frac{\eta^{2}}{2m}%
(\partial_{a}\phi)^{2}+V-\frac{\mu\eta^{2}}{2m^{2}}\frac{\nabla^{2}\rho^{1/2}%
}{\rho^{1/2}}\right]  ~.\label{energy d}%
\end{equation}
Requiring that $\dot{E}=0$ for arbitrary choices of $\dot{\rho}$ [which
follows from arbitrary choices of $\rho$ and $\phi$ in eq.(\ref{SEa})] we get
\begin{equation}
\eta\dot{\phi}+\frac{\eta^{2}}{2m}(\partial_{a}\phi)^{2}+V-\frac{\mu\eta^{2}%
}{2m^{2}}\frac{\nabla^{2}\rho^{1/2}}{\rho^{1/2}}=0~.\label{SEb}%
\end{equation}
Equations (\ref{SEa}) and (\ref{SEb}) are the coupled dynamical equations we
seek. The evolution of $\rho(x,t)$ in eq.(\ref{SEa}) is determined by
$\phi(x,t)$; the evolution of $\phi(x,t)$ in eq.(\ref{SEb}), is determined by
$\rho(x,t)$. The evolving geometry of the manifold enters through $\phi(x,t) $.

Next we show that, with one very interesting twist, the dynamical equations
turn out to be equivalent to the Schroedinger equation. We can always combine
the functions $\rho$ and $\phi$ into a complex function $\Psi=\rho^{1/2}%
\exp(i\phi)$. Then eqs.(\ref{SEa}) and (\ref{SEb}) can be rewritten as
\begin{equation}
i\eta\dot{\Psi}=-\frac{\eta^{2}}{2m}\nabla^{2}\Psi+V\Psi+\frac{\eta^{2}}%
{2m}\left(  1-\frac{\mu}{m}\right)  \frac{\nabla^{2}(\Psi\Psi^{\ast})^{1/2}%
}{(\Psi\Psi^{\ast})^{1/2}}\Psi~.\label{SEc}%
\end{equation}
This reproduces the Schroedinger equation,
\begin{equation}
i\hbar\frac{\partial\Psi}{\partial t}=-\frac{\hbar^{2}}{2m}\nabla^{2}%
\Psi+V\Psi~,\label{SE}%
\end{equation}
provided the osmotic mass is identified with the mass, $\mu=m$, and $\eta$ is
identified with Planck's constant, $\eta=\hbar$. Setting $S_{J}=\eta\phi$ in
eq.(\ref{SEb}) and letting $\eta\rightarrow0$ leads to the Hamilton-Jacobi
equation and thus the classical limit $\eta=\hbar\rightarrow0$ allows one to
identify $m$ with the inertial mass of the particle.

But why should $\mu=m$? This question is so central that we devote the next
section to it. But before that we note that the non-linearity in
eq.(\ref{SEc}) is undesirable both for experimental and theoretical reasons.
From the experimental side non-linear terms have been ruled out to an extreme
degree through precision experiments on the Lamb shift \cite{Smolin 86a} and
even more so in hyperfine transitions \cite{Bollinger 89}. From the theory
side there is a consistency argument that links the linearity of the Hilbert
space with the linearity of time evolution; retaining one and not the other
leads to inconsistently assigned amplitudes \cite{Caticha 98}. And, further,
it has been argued that the non-linear terms can lead to superluminal
communication \cite{Gisin 90}. Therefore it is extremely probable that the
identity of inertial and osmotic mass is exact.

Among the many mysteries of quantum theory there is one --- the central role
played by complex numbers --- that turns out to be related to these issues.
The dynamical equations (\ref{SEa}) and (\ref{SEb}) contain no complex numbers
but they can always be written in terms $\Psi$ and $\Psi^{\ast}$ instead of
$\rho$ and $\phi$. There is no mystery there. The statement that complex
numbers play a fundamental role in quantum theory is the non-trivial assertion
that the equation of evolution contains \emph{only} $\Psi$ and not \emph{both}
$\Psi$ \emph{and }$\Psi^{\ast}$. In the entropic approach both the linear time
evolution and the special role of complex numbers are linked through the
equality $\mu=m$.

\section{A new equivalence principle}

The generalization to many particles is easy. The conserved energy (for $N=2$)
[see eq.(\ref{gamma AB})] is%

\begin{align}
E  & =\int d^{3N}x\,\rho(x,t)\left(  A\gamma_{AB}v^{A}v^{B}+B\gamma_{AB}%
u^{A}u^{B}+V(x)\right)  \nonumber\\
& =\int d^{6}x\,\rho(x,t)\left(  \frac{1}{2}m_{1}v_{1}^{2}+\frac{1}{2}%
m_{2}v_{2}^{2}+\frac{1}{2}\mu_{1}u_{1}^{2}+\frac{1}{2}\mu_{2}u_{2}%
^{2}+V(x)\right)  ,
\end{align}
where we introduced inertial and osmotic masses, $m_{n}=2A/\sigma_{n}^{2}$ and
$\mu_{n}=2B/\sigma_{n}^{2}$. Note that the ratio of osmotic to inertial mass
turns out to be a universal constant, the same for all particles: $\mu
_{n}/m_{n}=B/A$. But why should $\mu_{n}=m_{n}$ \emph{exactly}? To see this
let us go back to eq.(\ref{energy c}). We can always change units and rescale
$\eta$ and $\tau$ by some constant $\kappa$ into $\eta=\kappa\eta^{\prime}$,
$\tau=\tau^{\prime}/\kappa$. If we also rescale $\phi$ into $\phi=\phi
^{\prime}/\kappa$, eqs.(\ref{SEa}) and (\ref{energy c}) become%
\begin{align}
\dot{\rho}  & =-\frac{\eta^{\prime}}{m}\left(  \partial^{a}\rho\partial
_{a}\phi^{\prime}+\rho\nabla^{2}\phi^{\prime}\right)  ~,\\
E  & =\int dx\,\rho\left(  \frac{\eta^{\prime2}}{2m}(\partial_{a}\phi^{\prime
})^{2}+\frac{\mu\kappa^{2}\eta^{\prime2}}{8m^{2}}(\partial_{a}\log\rho
)^{2}+V\right)  .
\end{align}
Then we can introduce a \emph{different }wave function $\Psi^{\prime}$ as
$\Psi^{\prime}=\rho^{1/2}\exp(i\phi^{\prime})$ which satisfies
\begin{equation}
i\eta^{\prime}\dot{\Psi}^{\prime}=-\frac{\eta^{\prime2}}{2m}\nabla^{2}%
\Psi^{\prime}+V\Psi^{\prime}+\frac{\eta^{\prime2}}{2m}\left(  1-\frac
{\mu\kappa^{2}}{m}\right)  \frac{\nabla^{2}(\Psi^{\prime}\Psi^{\prime\ast
})^{1/2}}{(\Psi^{\prime}\Psi^{\prime\ast})^{1/2}}\Psi^{\prime}.
\end{equation}
Since the mere rescaling by $\kappa$ can have no physical implications the
different regraduated theories are all equivalent and it is only natural to
use the simplest one: choose $\kappa=(A/B)^{1/2}$ so that $\mu\kappa^{2}=m$
and rescale the old $\mu$ to a new osmotic mass $\mu^{\prime}=\mu\kappa^{2}%
=m$. We conclude that \emph{whatever the value of the original coefficient
}$\mu$\emph{\ it is always possible to regraduate }$\eta$\emph{, }$\phi
$\emph{\ and }$\mu$\emph{\ to an equivalent but more convenient description
where the Schroedinger equation is linear and complex numbers attain a special
significance}. It is the rescaled value $\eta^{\prime}$ of the linear theory
that gets numerically identified with Planck's constant $\hbar$. From this
perspective the linear superposition principle and the complex Hilbert spaces
are important because they are convenient, but not because they are
fundamental --- a theme that was also explored in \cite{Caticha 98}.

These considerations remind us of Einstein's original argument for the
equivalence principle: We accept the complete physical equivalence of a
gravitational field with the corresponding acceleration of the reference frame
because this offers a natural explanation of the equality of inertial and
gravitational masses and opens the door to an explanation of gravity in purely
geometrical terms.

Similarly, in the quantum case \emph{we accept the complete equivalence of
quantum and statistical fluctuations because this offers a natural explanation
of the Schroedinger equation --- its linearity, its unitarity, the role of
complex numbers, the equality of intertial and osmotic masses --- and opens
the door to explaining quantum theory as an example of entropic inference.}

\section{Conclusions}

\noindent\textbf{On epistemology \emph{vs.} ontology:} Quantum theory has been
derived as an example of entropic dynamics. The discussion is explicitly
epistemological --- it is concerned with how we handle information and update
probabilities. Of course, once we know $x$ we can immediately make inferences
about the unobserved \textquotedblleft true\textquotedblright\ position $y$.
But this is precisely the point:\ the relation between the \textquotedblleft
laws of physics\textquotedblright\ (the Schroedinger equation) and
\textquotedblleft actual reality\textquotedblright\ is less direct than it is
commonly assumed.

\noindent\textbf{On interpretation:} Ever since Born the magnitude of the wave
function $|\Psi|^{2}=\rho$ has received a statistical interpretation. Within
the entropic dynamics approach the phase of the wave function is also
recognized as a feature of purely statistical origin. We can make this
explicit using eq.(\ref{curr}) to write the wave function as%
\begin{equation}
\Psi=\rho^{\frac{1+i}{2}}\,e^{iS}%
\end{equation}
where $\rho$ is a probability density and $S$ is an entropy --- it is the
entropy associated to each point on the statistical manifold $\mathcal{M}$.

\noindent\textbf{On dynamical laws:} The principles of entropic inference form
the backbone of this approach to dynamics. Energy conservation was introduced
as an important constraint in the present non-relativistic theory. One can
safely expect that in a fully relativistic theory it will not survive in its
current form. The peculiar features of quantum mechanics such as non-locality
and entanglement arise naturally by virtue of the theory being formulated in
the $3N$-dimensional configuration space.

\noindent\textbf{On time:} Time was introduced to keep track of the
accumulation of small changes and its particular form was chosen to simplify
the description of evolution. We have proposed a scheme that models temporal
order, its duration, and most interestingly, its directionality.

\noindent\textbf{Equivalence principle:} The derivation of the Schroedinger
equation from entropic inference led to a surprising similarity with general
relativity. The statistical manifold $\mathcal{M}$ is not a fixed background
but actively participates in the dynamics. The potential for uncovering deeper
relations between quantum theory and gravitation theory looks promising.

\noindent\textbf{Acknowledgments:} I would like to thank C. Rodr\'{\i}guez and
N. Caticha for their insightful advice, and D. Bartolomeo for pointing out an
important mistake in an earlier version of this paper. My deep appreciation
also goes to C. Cafaro, A. Giffin, P. Goyal, D. Johnson, K. Knuth, S. Nawaz,
and C.-Y. Tseng for many discussions, comments and suggestions.

\end{document}